\documentclass[article,  twocolumn,showpacs,preprintnumbers,amsmath,amssymb]{revtex4}
\usepackage{mathrsfs}

\usepackage{graphicx}
\usepackage{dcolumn}
\usepackage{bm}

\begin{document}
\author{Xiaohua Wu} \author{Bo You}
\address{Department of Physics, Sichuan University, Chengdu 610064, China.}
\title{ Unitary Application of the Quantum Error Correction Codes}
\begin{abstract}
From the set of operators for errors and its correction code, we
introduce the so-called complete unitary transformation. It can be
used for encoding while the inverse of it can be applied for
correcting the errors of the encoded qubit. We show that this
unitary protocol can be applied  for any code which satisfies the
quantum error correction condition.

\end{abstract}
 \maketitle In quantum computation and
communication, quantum error correction (QEC) will be necessary for
preserving coherent states against noise and other unwanted
interaction. Based on the classic schemes using redundancy, Shor [1]
has championed a strategy where a bit of quantum information is
stored in an entanglement of nine qubits. This scheme permits one to
correct for any error incurred by any of the nine qubits. For the
same purpose, Steane [2] has proposed a protocol which uses seven
qubits. Five qubit has the minimum size for a quantum code which
encodes a single qubit so that any error on a single qubit in the
encoded state can be detected and recovered. The five qubit code was
discovered by Bennett, DiVincenzo, Smolin and Wootters [3], and
independently by Laflamme, Miquel, Paz and Zurek [4]. The quantum
error-correction conditions were proved independently by Bennett and
co-authors [3], and by Knill and Laflamme [5].

The above protocols with different quantum error correction codes
(QECCs) can be viewed as active error correction. There are passive
error avoiding techniques such as the decoherence-free subspaces
[6-8] and noiseless subsystem [9-11]. Recently, it was found that
all the active and passive QEC methods can be unified
together[12-14].

The standard way of applying the known quantum error-correcting
codes (QECCs) for error-correcting contains: encoding procedure
$\mathcal{C}$, the noise channel $\mathcal{\varepsilon}$, and the
recovery operation $\mathcal{R}$. Considering the joint  system
$A\otimes B$, where $\{\vert e_i\rangle\}_{i=0,1,...,M}$ is the
basis of the ancilla system A while $\{\vert 0\rangle,\vert
1\rangle\}$ is the basis of the principle system B, the encoding
procedure can be realized with an unitary transformation U,
\begin{equation}
U\vert e_0\rangle\otimes \vert 0\rangle\rightarrow\vert 0_L\rangle,
U\vert e_0\rangle\otimes \vert 1\rangle\rightarrow\vert 1_L\rangle.
\end{equation}
Let $\rho^{\texttt{in}}$ denote the input state,
\begin{equation}
\rho^{\texttt{in}}=\vert e_0\rangle\langle
e_0\vert\otimes\vert\psi\rangle\langle\psi\vert,
\vert\psi\rangle=\alpha\vert 0\rangle+\beta\vert 1\rangle,
\end{equation}
after the operations of $\varepsilon$ and $\mathcal{R}$, the output
state $\rho^{\texttt{out}}$ is known,
\begin{equation}
\rho^{\texttt{out}}=(\mathcal{R}\circ\varepsilon)(U\rho^{\texttt{in}}U^{\dagger})=\vert
\Phi\rangle\langle\Phi\vert,
\end{equation}
where $\vert\Phi\rangle=\alpha\vert 0_L\rangle+\beta\vert
1_L\rangle.$ This standard QEC protocol is usually non-unitary: the
recovery operation $\mathcal{R}$ should  transfer the mixture
$\varepsilon(U\rho^{\texttt{in}}U^{\dagger})$ into the pure state
$\vert\Phi\rangle\langle\Phi\vert$. A different but unitary scheme
has been presented by Laflamme and co-authors.  They designed  a
five-qubit code and showed that the errors of the encoded qubit can
be corrected with a series of unitary transformations [4].

In the present work, we shall develop an unitary protocol to apply
the known perfect codes for quantum error correction. We introduce
the concept of complete unitary transformation $\tilde{U}$ which can
be decided by the code and the set of operators for errors. In the
unitary QEC protocol, $\tilde{U}$ is used for encoding while its
inverse $\tilde{U}^{\dagger}$ is sufficient  for correcting the
errors of the encoded qubit. Compared with the standard QEC
protocol, it leaves the errors of the ancilla system to be
un-corrected. The content of our work can be divided into three
parts. At first, we shall give a brief review for the work of
Laflamme and co-authors in [4], and generalize their work into the
unitary protocol where $\tilde{U}$ works. Then, we find a general
method to introduce $\tilde{U}$ and show that the unitary QEC
protocol, which is originated from the scheme in [4], can be applied
for any code satisfying the quantum error correction condition.
Finally, we show that our protocol is consistent with the unified
model of QEC developed by Kribs, Laflamme and Paulin in [12].

 To protect a qubit of information against the general one qubit errors,
Laflamme and co-authors presented the following five-qubits code,
\begin{eqnarray}
\vert 0_L\rangle=&&-\vert 00000\rangle+\vert 01111\rangle-\vert
10011\rangle+\vert 11100\rangle\nonumber\\
&&+\vert 00110\rangle+\vert 01001\rangle+\vert
10101\rangle+\vert 11010\rangle,\nonumber\\
\vert 1_L\rangle=&&-\vert 11111\rangle+\vert 10000\rangle+\vert
01100\rangle -\vert 00011\rangle\nonumber\\
&&+\vert 11001\rangle +\vert 10110\rangle-\vert 01010\rangle-\vert
00101\rangle.
\end{eqnarray}
They designed the quantum circuit for encoding and used the same
circuit running backwards for error-correcting. Their scheme is
organized  in Fig. 1a. Let the operators of errors are denoted by
$\varepsilon:\{\sqrt{p_m}E_m\}_{m=0.1,...,M}$, with $\langle
\Phi\vert E^{\dagger}_mE_m\vert\Phi\rangle=1$, the U in Fig. 1a has
the property that
\begin{eqnarray}
(\texttt{a})~~~ U\vert e_0\rangle\otimes \vert\psi\rangle
\rightarrow
\alpha\vert 0_L\rangle+\beta\vert 1_L\rangle,\nonumber \\
(\texttt{b})~~~ U^{\dagger}E_mU\vert e_0\rangle\otimes
\vert\psi\rangle=\vert e_m\rangle\otimes \vert
\psi_m\rangle,\nonumber
\end{eqnarray}
where the state $\vert \psi_m\rangle $ is known,
\[\vert\psi_m\rangle\in\{\pm(\alpha\vert 0\rangle+\beta \vert 1\rangle),
\beta\vert 0\rangle\pm\alpha\vert 1\rangle,\pm(\alpha\vert
0\rangle-\beta \vert 1\rangle)\}.\] Usually, we fix
$E_0=\texttt{\textbf{I}}$, and there should be
 $\vert\psi_0\rangle=\vert \psi\rangle$. The scheme in Fig. 1a works
 in the way like
\[\rho^{\texttt{out}}=U^{\dagger}[\varepsilon(U\rho^{\texttt{in}}U^{\dagger})]U=
 \sum_{m=0}^{M}p_m\vert e_m\rangle\langle e_m\vert\otimes\vert\psi_m\rangle\langle\psi_m\vert.\]
From it, the original state of the principle system can then be
restored by the successive unitary transformation
$U_{\delta}^{\dagger}$,
\begin{equation}
U_{\delta}^{\dagger}\vert e_m\rangle\otimes\vert
\psi_m\rangle\rightarrow \vert e_m\rangle\otimes\vert
\psi\rangle.\nonumber
\end{equation}
This $U_{\delta}^{\dagger}$ has been suggested in the original work,
the circuit for it has not been given there. As we shall show later,
it can be easily designed.

\begin{figure} \centering
\includegraphics[scale=0.50]{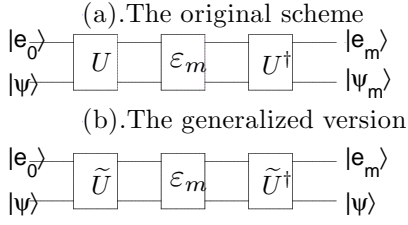}
\caption{\label{fig:epsart} (a) The scheme of the original work in
[4]. $U^{\dagger}$ is called the error finder there, it is realized
by the same circuit of U running backwards. (b) For the five qubit
code in (4),  we define $\tilde{U}=U\cdot U_{\delta}$ with
$U_{\delta}$ defined as $U^{\dagger}_{\delta}\vert
e_m\rangle\otimes\vert\psi_m\rangle\rightarrow\vert e_m\rangle
\otimes \vert\psi\rangle.$ Noting $U_{\delta}$ has been suggested in
[4] but its circuit was not given there. For other perfect codes,
the $\tilde{U}$ can be introduced by the general method in (11)}
\end{figure}

Jointing the two unitary $U$ and $U_{\delta}$ together, we could
define the complete unitary transformation $\tilde{U}$,
\begin{equation}
\tilde{U}=U\cdot U_{\delta},
\tilde{U}^{\dagger}=U_{\delta}^{\dagger}\cdot U^{\dagger}.
\end{equation}
 Noting $U_{\delta}^{\dagger}\vert
e_m\rangle\otimes\vert\psi_m\rangle\rightarrow\vert
e_m\rangle\otimes\vert\psi\rangle$, with
$U_{\delta}U_{{\delta}}^{\dagger}=\texttt{\textbf{I}}$, there should
be $U_{\delta}\vert
e_m\rangle\otimes\vert\psi\rangle\rightarrow\vert
e_m\rangle\otimes\vert\psi_m\rangle$. Jointing it with the known
property of U, one may easily verify that $\tilde{U}$ has the
following two properties:
\begin{eqnarray}
 \tilde{U}\vert e_0\rangle\otimes \vert\psi\rangle
\rightarrow
\alpha\vert 0_L\rangle+\beta\vert 1_L\rangle, \\
 \tilde{U}^{\dagger}E_m\tilde{U}\vert
e_0\rangle\otimes \vert\psi\rangle=\vert e_m\rangle\otimes \vert
\psi\rangle.
\end{eqnarray}
The result in (6) shows that $\tilde{U}$ can be used for encoding
and the one in (7) permits us to correct the errors of the encoded
qubit with $\tilde{U}^{\dagger}$. All these results are depicted in
fig. 1b where the total process can be described with
\begin{equation}
\rho^{\texttt{out}}=\tilde{U}^{\dagger}[\varepsilon(\tilde{U}\rho^{\texttt{in}}\tilde{U}^{\dagger})]\tilde{U}=
 \sum_{m=0}^{M}p_m\vert e_m\rangle\langle
 e_m\vert\otimes\vert \psi\rangle\langle\psi\vert.
 \end{equation}
 Compared with the standard QEC protocol, the errors of the ancilla system are not corrected here.

As a key step to show that the unitary protocol in Fig. 1b can be
applied for other perfect codes, we note that the  way of
introducing $\tilde{U}$ is non-unique. Besides the way in (5), we
find it can be also decided by the code in (4) and the operators of
errors. Let's introduce the denotation,
\begin{equation}
\vert 0,+\rangle\equiv\vert 0_L\rangle, \vert 0,-\rangle\equiv \vert
1_L\rangle,
\end{equation}
and define
\begin{equation}
E_m\vert 0,+\rangle=\vert m,+\rangle, E_m\vert 0,-\rangle=\vert
m,-\rangle. \end{equation} An interpretation for our denotation
above is shown in Fig. 2.
 With the code in (4) and the known
sixteen operators of errors, one may prove that the set of states,
$\{\vert m,\pm\rangle\}_{m=0,1,...,15}$, form an orthogonal basis.
Furthermore, one may also verify that the complete $\tilde{U}$ in
(5) is just the unitary transformation between the two sets of
basis, $\{\vert e_m\rangle\otimes \vert 0\rangle,\vert
e_m\rangle\otimes \vert 1\rangle\}_m$ and $\{\vert
m,\pm\rangle\}_m$, here,
\begin{equation}
\tilde{U}\vert e_m\rangle\otimes\left(
                          \begin{array}{c}
                            \vert 0\rangle \\
                            \vert 1\rangle \\
                          \end{array}
                        \right)\rightarrow\left(
                                            \begin{array}{c}
                                              \vert m,+\rangle \\
                                              \vert m.-\rangle \\
                                            \end{array}
                                          \right).
                      \end{equation}
Under the unitary condition that
$\tilde{U}\tilde{U}^{\dagger}=\texttt{\textbf{I}}$, there should be
\begin{equation}
\tilde{U}^{\dagger}\left(\begin{array}{c}\vert m,+\rangle \\
                                              \vert m.-\rangle \\
                                            \end{array}
                                          \right)\rightarrow\vert e_m\rangle\otimes\left(
                          \begin{array}{c}
                            \vert 0\rangle \\
                            \vert 1\rangle \\
                          \end{array}
                        \right).                    \end{equation}
\begin{figure} \centering
\includegraphics[scale=0.3]{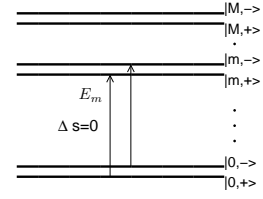}
\caption{\label{fig:epsart} The atomic model for QEC. We use $\vert
m.s\rangle$ to denote the level of the atom where m is the integer
for energy while s is the number of spin, $s=\pm 1$. Taking $\vert
0,\pm\rangle$ for the ground state,  it will be transited to the
$m-th$ level under the action of $E_m$. In this picture, the qubit
of information is stored in the internal degree of spin and this
information is protected since that all the transitions should obey
the rule $\Delta s=0$. }
\end{figure}

 The way of introducing $\tilde{U}$ in (11) is obviously general: For a given code and its
 corresponding set of errors $\{\sqrt{p_m}E_m\}_{m=0,1,...,M}$, we
 can  always introduce the set of states, $\{\vert
 m,\pm\rangle\}_{m=0,1,...,M}$, by following the steps in (9) and  (10). This set
 of states should formulate an orthogonal basis, as we shall show later, if the code satisfies the
 quantum error correction condition. Noting the basis,
  $\{\vert e_m\rangle\otimes\vert 0\rangle,\vert e_m\rangle\otimes\vert 1\rangle\}$, has also been given.
  In principle, one may get $\tilde{U}$ from (11) and design the quantum circuit  for it.
In following, we shall organize the above argument with a strict
proof: For any code which satisfies the perfect error-correcting
condition
 \begin{equation}
 \hat{P}_CE^{\dagger}_mE_n\hat{P}_c=\delta_{mn}\hat{P}_C
 \end{equation}
 where the projection operator $\hat{P}_C$ is defined as
$\hat{P}_C=\vert 0,+\rangle\langle 0,+\vert+ \vert 0,-\rangle\langle
 0,-\vert$, we have
$\hat{P}_CE^{\dagger}_mE_n\hat{P}_C=(\vert 0,+\rangle\langle
m,+\vert+\vert 0,-\rangle\langle m,-\vert)(\vert n,+\rangle\langle
0,+\vert+\vert n,-\rangle\langle 0,-\vert)$. Introducing the
following four Hermitian operators, $\hat{O}_1=\vert
0,+\rangle\langle 0,+\vert$, $\hat{O}_2=\vert 0,-\rangle\langle
0,-\vert$, $\hat{O}_3=\vert 0,+\rangle\langle 0,-\vert+\vert
0,-\rangle\langle 0,+\vert$, and $\hat{O}_4=i\vert 0,+\rangle\langle
0,-\vert-i\vert 0,-\rangle\langle 0,+\vert$, we could perform the
four calculations $\texttt{Tr}[\hat{O}_i(\cdot)]$ on the both sides
of equation (13) and get the results,
\begin{eqnarray}
\langle m,+\vert n,+\rangle=\langle m,-\vert
n,-\rangle=\delta_{mn},\nonumber\\
\langle m,+\vert n,-\rangle=\langle m,-\vert n,+\rangle=0,
\end{eqnarray}
which are sufficient to show that the set of states $\{\vert
 m,\pm\rangle\}_{m=0,1,...,M}$ formulate an orthogonal basis.
 With the $\tilde{U}$ from (11),
 we are able to show that the general scheme in Fig. 1b
 works for any perfect code. First, with $\vert
 \psi\rangle=\alpha\vert 0\rangle+\beta\vert 1\rangle$ and equation
 (11), we recover the result in (6), $\tilde{U}\vert e_0\rangle\otimes
 \vert\psi\rangle\rightarrow\alpha\vert 0,+\rangle+\beta\vert
 0,-\rangle$. Suppose that the error $E_m$ happens, from the denotation in (10),
 there is $E_m\tilde{U}\vert e_0\rangle=\alpha\vert
 m,+\rangle+\beta\vert m,-\rangle$. After the action of $\tilde{U}^{\dagger}$
 in (12), we have $\tilde{U}^{\dagger}E_m\tilde{U}\vert
 e_0\rangle=\vert e_m\rangle\otimes \vert\psi\rangle$, the same result given in (7).
Noting that the conditions in (6) and (7) are sufficient for
error-correcting of the principle system B, we conclude that any
perfect QECCs can be applied for error correction in the unitary way
shown in Fig. 1b.

\begin{figure} \centering
\includegraphics[scale=0.6]{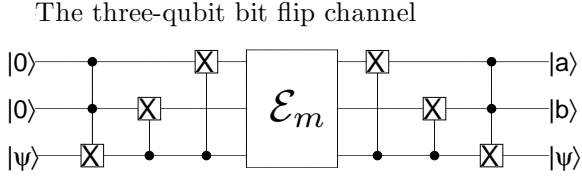}
\caption{\label{fig:epsart} The circuit for the three binary flip
cannel in [15]. Noting the circuit for encoding and the circuit of
error-correcting have a mirror symmetry.}
\end{figure}

It should be noted that $\tilde{U}$ is not unique. This can be seen
from the three qubit bit flip channel in [15]. Letting $\vert
e_0\rangle=\vert 00\rangle$, $\vert e_1\rangle=\vert 01\rangle$,
$\vert e_2\rangle=\vert 10\rangle$, $\vert e_3\rangle=\vert
11\rangle$, and fixing $E_0=I^{\otimes3}$, we still have the freedom
in defining the sequence of the operators. For example, the
following two choices, (I) $E_1=\hat{X}\otimes
\texttt{\textbf{I}}\otimes \texttt{\textbf{I}}$,
$E_2=\textbf{\texttt{I}}\otimes\ \hat{X}\otimes
\texttt{\textbf{I}}$, $E_3=\texttt{\textbf{I}}\otimes
\textbf{\texttt{I}}\otimes\ \hat{X}$ and (II)
$E_1=\textbf{\texttt{I}}\otimes\ \hat{X}\otimes
\texttt{\textbf{I}}$, $E_2=\hat{X}\otimes \texttt{\textbf{I}}\otimes
\texttt{\textbf{I}}$, $E_3=\texttt{\textbf{I}}\otimes
\textbf{\texttt{I}}\otimes\ \hat{X}$, will lead two different
$\tilde{U}$ which can  both be applied for Fig. 1b. However, the
circuits for them are different. So, the sequence of the operators
should be specified when the quantum circuit for $\tilde{U}$ is to
be designed. The circuit in Fig. 3 is for the three-qubit bit flip
channel with $\vert 0_L\rangle =\vert 000\rangle$, $\vert
1_L\rangle=\vert 111\rangle$, and the sequence of the operators in
(I) above. The circuit in Fig. 4 is constructed for the five-qubit
code,
\begin{eqnarray}
\vert 0_L\rangle&=&-\vert 00000\rangle+\vert 00101\rangle+\vert
01010\rangle+\vert 01111\rangle\nonumber\\
&+&\vert 10011\rangle-\vert 10110\rangle+\vert 11001\rangle+\vert
11100\rangle\nonumber\\
\vert 1_L\rangle&=&-\vert 00011\rangle-\vert 00110\rangle+\vert
01001\rangle-\vert 01100\rangle\nonumber\\
&+&\vert 10000\rangle+\vert 10101\rangle+\vert 11010\rangle-\vert
11111\rangle,
\end{eqnarray}
which is get from the code in (4) by moving the third qubit to the
final location. The sequence of the operators is:
\[\hat{I},\hat{X}_4,\hat{Z}_3,\hat{X}_5,\hat {Z}_2,
\hat{Y}_3,\hat{X}_1,\hat{X}_3,
\hat{Z}_1,\hat{Y}_5,\hat{Z}_5,\hat{X}_2,\hat{Z}_4,\hat{Y}_4,
\hat{Y}_1,\hat{Y}_2,\] while the basis vectors $\vert e_m\rangle$
are fixed as $\vert e_0\rangle=\vert 0000\rangle$, $\vert
e_1\rangle=\vert 0001\rangle$, $\vert e_2\rangle=\vert
0010\rangle$,..., $\vert e_{15}\rangle=\vert 1111\rangle$.

\begin{figure} \centering
\includegraphics[scale=0.6]{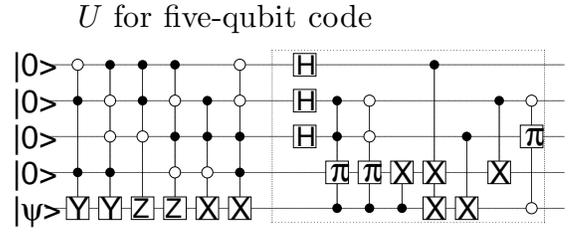}
\caption{\label{fig:epsart} In the original circuit for the five
qubit code in (4), the information is encoded in the third qubit. In
the present work, we use the code in (15) and encode the qubit of
information in the final location. The part of circuit, which is
within the dash lines,
 plays the role of $U$ in Fig. 1a. It is designed in the similar way of [4].
H is used for the Hadamard gate. The filled circle denotes the
control is $\vert 1\rangle$ while the empty one is for $\vert
0\rangle$. $\pi$ is  the global phase shift $\exp\{i\pi\}$ in short.
$\tilde{U}^{\dagger}$ is not given here, it can be easily
constructed by letting the above circuit run backwards.
  }
\end{figure}

Considering the fact that both the $\tilde{U}$ and
$\varepsilon:\{\sqrt{p_m}E_m\}$ are known, we could introduce the
so-called transformed operators,
\begin{equation}
\tilde{E}_m=\tilde{U}^{\dagger}E_m\tilde{U},
\end{equation}
and define the transformed channel as
$\tilde{\varepsilon}:\{\sqrt{p_m}\tilde{E}_m\}$ with $\langle
\Phi\vert \tilde{E}_m^{\dagger}\tilde{E}_m\vert\Phi\rangle=1.$
Certainly, there should be
\begin{equation}
\tilde{E}_m\vert e_0\rangle\otimes\vert \psi\rangle=\vert
e_m\rangle\otimes \vert\psi\rangle.
\end{equation}
Now, the process in Fig. 1b can be expressed with the compact form
\begin{equation}
\rho^{\texttt{out}}=\tilde{\varepsilon
}(\rho^{\texttt{in}})=\sum_{m=0}^{M}p_m\tilde{E}_m\rho^{\texttt{in}}\tilde{E}_m^{\dagger}.
\end{equation}
Certainly, $\rho^{\texttt{out}}=\sum_{m=0}^M p_m\vert
e_m\rangle\langle e_m\vert\otimes \vert\psi\rangle\langle\psi\vert$.
As it is shown in [12], the QEC with perfect codes can be unified
with other QEC protocols like the decoherence-free subspaces and the
noiseless subsystems. The unified scheme for quantum
error-correction consists of a triple $(\mathcal{R},
\varepsilon,\mathscr{U})$, $\mathscr{U}$ is correctable for
$\varepsilon$ if
\begin{equation}
(\texttt{Tr}_\texttt{A}\circ\mathcal{P}_{\mathscr{U}}\circ\mathcal{R}\circ\varepsilon)(\rho)=\texttt{Tr}_{\texttt{A}}(\rho).
\end{equation}
It can be shown that $\tilde{\varepsilon}$ is consistent with this
unified scheme. At first, we introduce the decomposition of the
joint system $A\otimes B$, $\mathcal
{H}=(\mathcal{H}^{\texttt{A}}\otimes
\mathcal{H}^\texttt{B})\oplus\mathcal {K}$, where the basis for each
subspace is known: $\mathcal{H}^{\texttt{A}}$ is one-dimensional
with $\vert e_0\rangle$, $\mathcal {H}^{\texttt{B}}$ is with its
basis as $\{\vert 0\rangle,\vert 1\rangle\}$, and $\mathcal {K}$ has
its basis to be $\{\vert e_m\rangle\otimes 0\rangle,\vert
e_m\rangle\otimes \vert 1\rangle\}$ for $m\ge1$. Then, we could
define a set of operators
\begin{equation}
 \mathscr{U}=\{\rho\in \mathcal {B}(\mathcal {H}), \rho=\vert e_0\rangle\langle e_0\vert\otimes \rho^{\texttt{B}}\}
 \end{equation}
where $\rho^{\texttt{B}}$ is an arbitrary state of the principle
system B. With $\hat{P}_{\mathscr{U}}=\vert e_0\rangle\langle
e_0\vert\otimes(\vert 0\rangle\langle 0\vert+\vert 1\rangle\langle
1\vert)$, we have $\hat{P}_{\mathscr{U}}\mathcal {H}=\mathcal
{H}^{\texttt{A}}\otimes\mathcal {H}^{\texttt{B}}$. Let $\mathcal
{P}_{\mathscr{U}}=\hat{P}_{\mathscr{U}}(\cdot)\hat{P}_{\mathscr{U}}$,
we find that our protocol in (18) could be expressed as
\begin{equation}
(\texttt{Tr}_\texttt{A}\circ\mathcal{P}_{\mathscr{U}}\circ\tilde{\varepsilon})(\rho)=\texttt{Tr}_{\texttt{A}}(\rho),\forall{\rho}\in\mathscr{U}.
\end{equation}
In other words, it is captured in the unified scheme with the
recovery operation $\mathcal {R}=\texttt{\textbf{I}}$.

For simplicity, we have expressed the operators of the errors with
the form $\{\sqrt{p_m}E_m\}$. This denotation is strict if the code
saturates  the quantum Hamming bound. For the more general case, one
may introduce an extra index besides the subscript m for the
operators, say , $E^{\alpha_m}_m$,  and let $\{E_m^{\alpha_m}\}$
denote the subset of the operators whose action on $\vert
0,\pm\rangle$ will lead to the same state, $E_m^{\alpha_m}\vert
0,\pm\rangle=\vert m,\pm\rangle$. This substitution, $E_m\rightarrow
E_m^{\alpha_m}$, will  not change the results above.

With a simple program, we have got the complete $\tilde{U}$
corresponding to the Shor's  nine qubit code, Steane's seven qubit
code, and the five qubit code of Bennett and co-authors. For each
$\tilde{U}$, we have calculated all the deformed Kraus operators,
 $\tilde{U}^{\dagger}E_m\tilde{U}$, and verified that the result in
 (16) always holds. The quantum circuit for these complete unitary transformation are still
 under researching.
 Suppose the designed circuit has been realized in experiment,
one could perform the standard quantum process tomography (SQPT)
over the channel of the encoded qubit [15].  With the experimental
data about the four final states of system B, which correspond to
the set of input states, $\vert e_0\rangle\otimes\vert\phi_j\rangle$
with $\forall\vert\phi_j\rangle\in\{\vert 0\rangle, \vert 1\rangle,
\frac{\sqrt{2}}{2}(\vert 0\rangle+\vert
1\rangle),\frac{\sqrt{2}}{2}(\vert 0\rangle+i\vert 1\rangle)\}$, one
may easily judge whether the channel of B is perfect or not.

 Compared with the standard  QEC
protocol, the scheme in Fig. 1b does not require the errors in the
ancilla system to be corrected. In some aspects, our scheme is very
similar with the passive QEC protocols where the  recovery operation
$\mathcal{R}$ takes a trivial form. As a known result, any code
satisfying the quantum error-correction condition in (13) can be
used in the standard QEC protocol. For the same code, we offer
another choice   of applying it for quantum error correction.

We would like to acknowledge the help discussion with Prof. Cen
L.-X.

\end{document}